\documentclass{article}

\newcommand{\p}[1]{(\ref{#1})}

\newcommand{\cR}{{\cal R}}
\newcommand{\tR}{{\textsf R}}
\newcommand{\tbR}{\overline{\textsf R}}
\newcommand{\tJ}{{\textsf J}}
\newcommand{\cbR}{\overline{\cal R}}
\newcommand{\cJ}{{\cal J}}
\newcommand{\hJ}{{\hat J}}

\newcommand{\bR}{{\overline R}{}}

\newcommand{\bpsi}{{\bar\psi}{}}

\newcommand{\bp}{{\bar p}}
\newcommand{\bz}{{\bar z}}

\newcommand{\be}{\begin{equation}}
\newcommand{\ee}{\end{equation}}
\newcommand{\bea}{\begin{eqnarray}}
\newcommand{\eea}{\end{eqnarray}}

\newcommand{\ba}{\begin{array}} \newcommand{\ea}{\end{array}}

\newcommand{\nn}{\nonumber}

\usepackage{amscd,amsmath,amssymb}

\begin{document}
\thispagestyle{empty}
\vspace{2cm}
\begin{flushright}
\end{flushright}\vspace{2cm}
\begin{center}
{\Large\bf Symmetries of $N=4$ supersymmetric $\mathbb{CP}^n$ mechanics}
\end{center}
\vspace{1cm}

\begin{center}
{\large\bf S.~Bellucci${}^a$, N.~Kozyrev${}^b$, S.~Krivonos${}^{b}$ and A.~Sutulin${}^{b}$ }
\end{center}

\begin{center}
${}^a$ {\it
INFN-Laboratori Nazionali di Frascati,
Via E. Fermi 40, 00044 Frascati, Italy} \vspace{0.2cm}

${}^b$ {\it
Bogoliubov  Laboratory of Theoretical Physics, JINR,
141980 Dubna, Russia}

\end{center}
\vspace{2cm}

\begin{abstract}\noindent
We explicitly constructed the generators of $SU(n+1)$ group which commute with the supercharges
of $N=4$ supersymmetric $\mathbb{CP}^n$ mechanics in the background $U(n)$ gauge fields. The corresponding classical Hamiltonian
can be represented as a direct sum of two Casimir operators: one Casimir operator on $SU(n+1)$ group contains
our bosonic and fermionic coordinates and momenta, while the second one, on the $SU(1,n)$ group, is  constructed from isospin degrees of freedom only.

\end{abstract}

\newpage
\setcounter{page}{1}
\setcounter{equation}{0}
\section{Introduction}
The construction of supersymmetric non-linear sigma-models in one
dimensions has been known for  many years \cite{BZ,DMPH,MP}. If
the scalar fields take values in a complex K\"{a}hler manifold,
then the supersymmetric Lagrangian is given by an elegant and
simple expression in terms of the K\"{a}hler metric. In the
simplest case of the supersymmetric $\mathbb{CP}^n$ model
\cite{CS} in one dimension the $N=4$ supercharges acquire, in the
appropriate basis, a very simple form \cite{sm1,BN1,BN2}. The main
properties of the model can be described by the statement that the
supercharges and the Hamiltonian are invariant under the action of
the $SU(n+1)$ group, whereas the bosonic component fields
parameterize the $SU(n+1)/U(n)$ manifold \cite{MacF1}.

One of the interesting further questions is how to introduce the
interaction in the supersymmetric $\mathbb{CP}^n$ model without
breaking the $SU(n+1)$ symmetry. Clearly, the evident guess is to
try to introduce the interaction with non-Abelian background
fields living on a $U(n)$ subgroup. In the bosonic case such a
system has been proposed by Karabali and Nair who constructed
higher dimensional quantum Hall systems on $\mathbb{CP}^n$
manifolds \cite{KN2}.\footnote{Non-compact versions have been
considered in \cite{ncQHE1, ncQHE2}.} The corresponding bulk and
edge actions were derived \cite{Kar1}. The supersymmetrization of
this construction is not a straightforward task because in many
cases adding the interaction with background fields results in the
so called "weak supersymmetry" algebra \cite{weak}.

The program to construct $N=4$ supersymmetric extensions of the
Karabali and Nair approach was initiated in \cite{BKS01}, where a
$N=4$ supersymmetric mechanics describing the motion of a charged
particle over the $\mathbb{CP}^n$ manifold in the presence of
background $U(n)$ gauge fields was constructed.  In contrast with
$N=4$ supersymmetric mechanics on $S^4$ \cite{KLS01} which
possesses only the $SO(4)$ invariance, the $N=4$ supersymmetric
mechanics on $\mathbb{CP}^n$ is invariant under the full $SU(n+1)$
group, which of course is realized non-linearly. Thus, the key
point is to understand whether the inclusion of non-Abelian $U(n)$
background gauge fields in $N=4$ supersymmetric mechanics on
$\mathbb{CP}^n$ preserves the $SU(n+1)$ symmetry of the model. In
the present paper we explicitly construct the currents, forming
the $su(n+1)$ algebra with respect to a standard Poisson brackets,
which commute with the $N=4$ supercharges of the model
\cite{BKS01}. We also reveal the nice structure of the classical
Hamiltonian of $N=4, \mathbb{CP}^n$ supersymmetric mechanics in
the presence of background $U(n)$ gauge fields to be a direct sum
of two Casimir operators: one Casimir operator on the $SU(n+1)$
group contains our bosonic and fermionic coordinates and momenta,
while the second one, on the $SU(1,n)$ group, is constructed from
isospin degrees of freedom only.

The paper is organized as follows. In Section 2 we review the
symmetry properties of the bosonic $\mathbb{CP}^n$ mechanics with
and without background Abelian and non-Abelian gauge fields. In
Section 3, after a short discussion of the symmetries of $N=4$
supersymmetric $\mathbb{CP}^n$ model in the absence of background
fields, we present the explicit form of the $su(n+1)$ currents
commuting with the supercharges. The summary of our results and
perspectives for future studying is collected in the Conclusion.

\setcounter{equation}{0}
\section{Preliminaries: Bosonic $\mathbb{CP}^n$ mechanics and its symmetries}
\subsection{Bosonic $\mathbb{CP}^n$ model: Lagrangian approach}
The standard bosonic $\mathbb{CP}^n$ model is defined in terms of $2n$ bosonic coordinates $\{z^\alpha, \bz_\alpha,\; \alpha=1,\ldots,n\}$ depending on time $t$ by the Lagrangian
\be\label{L}
L =\int dt {\cal L} = \int dt \; g_\alpha{}^\beta\; {\dot z}{}^\alpha \, {\dot \bz}_\beta\,,
\ee
where the $\mathbb{CP}^n$ metric $g_\alpha{}^\beta$ has the standard Fubini-Study form
\be\label{metrics}
g_\alpha{}^\beta=\frac{1}{\left( 1+ z\cdot \bz\right)}\left[\delta_\alpha^\beta -\frac{\bz_\alpha z^\beta}{\left( 1+ z\cdot \bz\right)}\right], \quad z\cdot \bz\equiv z^\alpha \bz_\alpha.
\ee
$\mathbb{CP}^n$ mechanics provides a very simple example of the non-linear realization of $SU(n+1)$ symmetry in which a $U(n)$ subgroup is realized linearly \cite{MacF1}. Thus, $\mathbb{CP}^n$ mechanics could be interpreted as the $\sigma$-model on the coset $SU(n+1)/U(n)$. The corresponding construction is quite simple.

Firstly, let us fix the commutation relations of the $su(n+1)$
algebra to be
\bea\label{sun1} && \left[
\tR_\alpha, \tbR^\beta\right] = 2\,\tJ_\alpha{}^\beta, \quad
\left[\tJ_\alpha{}^\beta, \tJ_\gamma{}^\sigma
\right]=\frac{1}{2}\left(
\delta_\gamma^\beta \tJ_\alpha{}^\sigma - \delta_\alpha^\sigma \tJ_\gamma{}^\beta\right), \nn \\
&& \left[ \tJ_\alpha{}^\beta,  \tR_\gamma \right] =\frac{1}{2}\left(\delta_\gamma^\beta \tR_\alpha+\delta_\alpha^\beta \tR_\gamma\right), \quad
\left[ \tJ_\alpha{}^\beta, \tbR^\gamma \right] = -\frac{1}{2}\left( \delta_\alpha^\gamma \tbR^\beta+\delta_\alpha^\beta \tbR^\gamma \right) .
\eea
Thus, the generators $ \tR_\alpha, \tbR^\alpha$ belong to the coset $SU(n+1)/U(n)$,
while the $\tJ_\alpha{}^\beta$ form $u(n)$ subalgebra. In addition,  these generators are chosen to be
hermitian ones
\be\label{anti}
\left( \tR_\alpha \right)^\dagger = \tbR^\alpha, \quad
\left( \tJ_\alpha{}^\beta\right)^\dagger= \tJ_\beta{}^\alpha.
\ee

Now, we can realize the action of $SU(n+1)$ group on the $SU(n+1)/U(n)$ coset element $g$
\be\label{gsu}
g=e^{i \left(x^\alpha\, \tR_\alpha +{\bar x}_\alpha \tbR^\alpha\right)}
\ee
by the left multiplications as
\be
g_0\;g = e^{i \left(a^\alpha\, \tR_\alpha +{\bar a}_\alpha \tbR^\alpha+ b_\alpha{}^\beta \tJ_\beta{}^\alpha\right)}\,e^{i \left(x^\alpha\, \tR_\alpha +{\bar x}_\alpha \tbR^\alpha\right)} = e^{i \left(x'{}^\alpha\, \tR_\alpha +{\bar x}'{}_\alpha \tbR^\alpha\right)}\, h,
\ee
where $h\in U(n)$. The explicit form of $\{ \tR_\alpha,\tbR{}^\alpha\}$ transformations reads
\be\label{tr0}
z'{}^\alpha=z^\alpha+ a^\alpha + (z\cdot {\bar a}) z^\alpha,\quad
{\bar z}'_{\alpha}={\bar z}_\alpha+{\bar a}_\alpha +(a\cdot {\bar z}) {\bar z}_\alpha,
\ee
where $\{z^\alpha, {\bar z}_\alpha\}$ are defined as
\be\label{tr1}
 z^\alpha \equiv \frac{ \tan \sqrt{ x \cdot {\bar x}}}{\sqrt{ x \cdot {\bar x}}}x^\alpha,\quad
 {\bar z}{}^\alpha \equiv \frac{ \tan \sqrt{ x \cdot {\bar x}}}{\sqrt{ x \cdot {\bar x}}}{\bar x}{}^\alpha.
\ee
One may easily check that the Lagrangian \p{L} is indeed invariant under transformations \p{tr0}.

The basic covariant objects in the coset approach are the Cartan forms
\be\label{CF}
g^{-1}\,d g = i\, dz^\alpha\, e_\alpha{}^\beta \tR_\beta +i\, \tbR^\alpha e_\alpha{}^\beta d\bz_\beta
+ 2  \left( z^\alpha \, \omega_\beta{}^\gamma d \bz_\gamma
- dz^\gamma\, \omega_\gamma{}^\alpha \bz_\beta \right)\, \tJ_\alpha{}^\beta.
\ee
With our definitions the explicit expressions for the vielbeins $e_\alpha{}^\beta$ and
$U(n)$-connections $\omega_\alpha{}^\beta$ on the $\mathbb{CP}^n$ manifold entering \p{CF} read \cite{klw}
\bea
&&
e_\alpha{}^\beta = \frac{1}{\sqrt{1+z\cdot \bz}}\left[ \delta_\alpha^\beta -
\frac{\bz_\alpha z^\beta}{\sqrt{1+z\cdot \bz}\left(1+\sqrt{1+z\cdot \bz}\right)}\right],
\label{vb} \\
&&
\omega_\alpha{}^\beta= \frac{1}{\sqrt{ 1+ z \cdot \bz
}\left(1+\sqrt{1+ z \cdot \bz}\right)}\left[ \delta_\alpha^\beta-
\frac{\bz_\alpha z^\beta}{2\,\sqrt{1+ z \cdot \bz}\left(1+\sqrt{1+ z \cdot \bz}\right)}\right].
\label{con}
\eea
The vielbeins $e_\alpha{}^\beta$ define the $SU(n+1)$ covariant derivatives of our "fields" $\{z^\alpha, {\bar z}_\alpha\}$
\be
\nabla_t z^\alpha = {\dot z}^\beta\, e_\beta{}^\alpha, \quad \nabla_t {\bar z}_\alpha = e_\alpha{}^\gamma {\dot \bz}_\gamma
\ee
and the $SU(n+1)$ invariant Lagrangian density unambiguously restored to be
\be\label{L1}
{\cal L}= \nabla_t z^\alpha\, \nabla_t {\bar z}_\alpha = g_\alpha{}^\beta\; {\dot z}{}^\alpha \, {\dot \bz}_\beta\,,
\ee
which coincides with \p{L}.

\subsection{Bosonic $\mathbb{CP}^n$ model: Hamiltonian approach}
The Lagrangian approach we considered in the previous Subsection
has a nice geometric interpretation within the coset construction.
In contrast, the Hamiltonian approach is more flexible and
provides a simple possibility to introduce the interactions with
the background gauge fields in the pure $\mathbb{CP}^n$
mechanics.
\subsubsection{$\mathbb{CP}^n$ model}
The  Hamiltonian of the $\mathbb{CP}^n$ models \p{L} reads
\be\label{bH}
H=\bp^\alpha\, \left( g^{-1}\right)_\alpha{}^\beta \, p_\beta \;,
\ee
with
\be
 \left(g^{-1}\right)_\alpha{}^\beta = \left( 1+ z\cdot \bz\right)\left[\delta_\alpha^\beta+\bz_\alpha z^\beta\right].
\ee

The $SU(n+1)$ invariance of the  $\mathbb{CP}^n$ model at the Hamiltonian level  means that the Hamiltonian  \p{bH} has vanishing Poisson brackets with the currents
\be\label{subos}
R_\alpha= p_\alpha +\bz_\alpha \bz_\beta \bp^\beta,\;
\bR{}^\alpha=\bp{}^\alpha+z^\alpha z^\beta p_\beta,\;
J_\alpha{}^\beta = \frac{i}{2}\left( z^\beta p_\alpha -\bz_\alpha \bp{}^\beta\right)+
\frac{i}{2}\delta_\alpha^\beta\left( z^\gamma p_\gamma -\bz_\gamma \bp{}^\gamma\right).
\ee
These currents form the $su(n+1)$ algebra
\bea\label{sunp1}
&& \left\{R_\alpha, \bR{}^\beta\right\}  = 2 i J_\alpha{}^\beta ,\quad
\left\{J_\alpha{}^\beta, J_\gamma{}^\delta\right\} = \frac{i}{2} \left( \delta_\gamma^\beta\; J_\alpha{}^\delta-\delta_\alpha^\delta\; J_\gamma{}^\beta\right),
\nn \\
&& \left\{J_\alpha{}^\beta, R{}_\gamma\right\} = \frac{i}{2} \left( \delta_\gamma^\beta\; R_\alpha+\delta_\alpha^\beta\; R_\gamma\right) , \quad
\left\{J_\alpha{}^\beta, \bR{}^\gamma\right\} = -\frac{i}{2} \left( \delta_\alpha^\gamma \; \bR{}^\beta+\delta_\alpha^\beta\; \bR{}^\gamma\right)
\eea
with respect to the standard Poisson brackets
\be\label{pb1}
\left\{ z^\alpha, p_\beta \right\} = \delta^\alpha_\beta, \qquad
\left\{ \bz_\alpha, \bp^\beta \right\} = \delta_\alpha^\beta\;.
\ee
Moreover, one may check that the Hamiltonian \p{bH} coincides with the quadratic $su(n+1)$ Casimir operator
\be\label{Cas}
{\cal C} = R_\alpha\; \bR{}^\alpha +2 J_\alpha{}^\beta\; J_\beta{}^\alpha -\frac{2}{n+1} J_\alpha{}^\alpha\; J_\beta{}^\beta ,
\ee
with the realization \p{subos} taken into account. This is just another visualization of the $SU(n+1)$ invariance of $\mathbb{CP}^n$ model.
\subsubsection{Constant magnetic field}
If we are interested in the introduction of some interaction in
our system which preserves the $SU(n+1)$ symmetry, then we should
think about the interaction with background magnetic fields only,
because it is impossible to construct any $SU(n+1)$ invariant
potential term from our coordinates $\{z^\alpha, {\bar
z}_\alpha\}$ which transform as in \p{tr0}. Moreover, in
accordance with the $SU(n+1)/U(n)$ interpretation of
$\mathbb{CP}^n$ mechanics, in order to preserve all symmetries,
the background fields could be either Abelian (which corresponds
to the $U(1)$ group in the stability subgroup $U(n)$) or
non-Abelian, living on the whole $U(n)$ subgroup.

The simplest way to introduce the interaction with the Abelian $U(1)$ background magnetic field $B$ is to modify the $U(n)$ currents $J_\alpha{}^\beta$ \p{subos} as
\be\label{newJ}
{\widetilde J}_\alpha{}^\beta = J_\alpha{}^\beta+B\;\delta_\alpha^\beta.
\ee
It is a matter of straightforward calculations to find new coset $SU(n+1)/U(n)$ currents $\{{\widetilde R}_\alpha, {\widetilde \bR}{}^\alpha\}$
\be\label{B1}
{\widetilde R}_\alpha = R_\alpha+i B\; \bz_\alpha, \qquad
{\widetilde \bR}{}^\alpha = \bR{}^\alpha-i B\; z^\alpha,
\ee
which form, together with ${\widetilde J}_\alpha{}^\beta$, the same $su(n+1)$ algebra \p{sunp1}.

If we now  define the Hamiltonian as  the Casimir operator \p{Cas} constructed from the generators \p{B1}, then  we will get
\be\label{BHam}
H={\overline{\widetilde p}}{}^\alpha\; \left( g^{-1}\right)_\alpha{}^\beta \;{\widetilde p}_\beta+
\frac{2 n}{n+1} B^2 ,
\ee
where
\be\label{Bmom}
\left\{
\begin{array}{l}
{\widetilde p}{}_\alpha = p_\alpha - i B\; \frac{\bz_\alpha}{1+z\cdot \bz}\\
\\
{\overline{\widetilde p}}{}^\alpha = \bp^\alpha + i B\; \frac{z^\alpha}{1+z\cdot \bz}\\
\end{array}
\right.\quad\Rightarrow\quad \left\{ {\tilde p}_\alpha,
{\overline{\tilde p}}{}^\beta\right\} = -2 i B g_\alpha{}^\beta .
\ee Clearly, the Hamiltonian \p{BHam} describes $\mathbb{CP}^n$
mechanics in the uniform $U(1)$ background.

One can include also some analog of the oscillator potential
fields, which naturally breaks the full $SU(n+1)$ symmetry down to
$U(n)$, but nonetheless preserves the exact solvability and all
$\mathbb{CP}^n$ oscillator symmetries, including hidden ones, even
in the presence of a constant magnetic field \cite{new1}.

\subsubsection{Non-Abelian magnetic field}
The same strategy we used in the previous Subsection can be applied for the more interesting case of non-Abelian $U(n)$ background.

Similarly to the case of $U(1)$ background, we will start with the modification of $U(n)$ currents $J_\alpha{}^\beta$ \p{subos} now as
\be\label{newJsu}
{\widetilde J}_\alpha{}^\beta  = J_\alpha{}^\beta+\hJ_\alpha{}^\beta.
\ee
Here, the currents $\hJ_\alpha{}^\beta$ have vanishing brackets with the coordinates and momenta $\{z^\alpha, {\bar z}_\alpha, p_\alpha,\bar{p}^\alpha\}$ and form the same $U(n)$ algebra \p{sunp1}
\be\label{inneru}
\left\{\hJ_\alpha{}^\beta, \hJ_\gamma{}^\delta\right\} = \frac{i}{2} \left( \delta_\gamma^\beta\; \hJ_\alpha{}^\delta-\delta_\alpha^\delta\; \hJ_\gamma{}^\beta\right).
\ee
Now, one has to construct the new $SU(n+1)/U(n)$ currents which will span, together with ${\widetilde J}_\alpha{}^\beta$ \p{newJsu}, the $su(n+1)$ algebra. After some calculations one may find these modified $su(n+1)$ generators
\bea\label{Ab1}
&& {\widetilde R}_\alpha = R_\alpha+\frac{2 i}{\left(1+\sqrt{1+ z \cdot \bz}\right)}\; \hJ_\alpha{}^\beta \bz_\beta +\frac{ i}{\left(1+\sqrt{1+ z \cdot \bz}\right)^2}\;\bz_\alpha \;z^\beta \hJ_\beta{}^\gamma \bz_\gamma ,\nn \\
&& \overline{\widetilde R}{}^\alpha = \bR{}^\alpha - \frac{2 i}{\left(1+\sqrt{1+ z \cdot \bz}\right)}\; z^\beta \hJ_\beta{}^\alpha{} -\frac{ i}{\left(1+\sqrt{1+ z \cdot \bz}\right)^2}\;z^\alpha \;z^\beta \hJ_\beta{}^\gamma \bz_\gamma ,\nn \\
&& {\widetilde J}_\alpha{}^\beta = -\frac{i}{2}\left\{ {\widetilde R}_\alpha, \overline{\widetilde R}{}^\beta \right\}=J_\alpha{}^\beta+\hJ_\alpha{}^\beta.
\eea
The corresponding Hamiltonian can be again defined as the Casimir operator \p{Cas} constructed now from the currents \p{Ab1}. Explicitly it reads
\be\label{AbHam}
H={\overline{\widetilde p}}{}^\alpha\; \left( g^{-1}\right)_\alpha{}^\beta \;{\widetilde p}_\beta+
2\; \hJ_\alpha{}^\beta \; \hJ_\beta{}^\alpha -\frac{2}{n+1} \; \hJ_\alpha{}^\alpha \; \hJ_\beta{}^\beta ,
\ee
where now
\be\label{Abmom}
\left\{
\begin{array}{l}
{\widetilde p}{}_\alpha = p_\alpha - 2 i\; \omega_\alpha{}^\beta \; \hJ_\beta{}^\gamma\;\bz_\gamma\\
\\
{\overline{\widetilde p}}{}^\alpha = \bp^\alpha + 2 i \; z^\gamma \; \hJ_\gamma{}^\beta\;\omega_\beta{}^\alpha\\
\end{array}
\right.\quad\Rightarrow\quad
\left\{ {\tilde p}_\alpha, {\overline{\tilde p}}{}^\beta\right\} =
- 2 \;i \; e_\alpha{}^\mu\; e_\nu{}^\beta \; \hJ_\mu{}^\nu,
\ee
with vielbeins $e_\alpha{}^\beta$ and $U(n)$ connections $\omega_\alpha{}^\beta$ on $SU(n+1)/U(n)$ defined in \p{vb}, \p{con}, respectively.

Let us note that the $\hJ_\alpha{}^\beta \rightarrow B \delta_\alpha^\beta$ reduction brings us from the non-Abelian to
the $U(1)$ Abelian case. Another comment concerns the structure of the Hamiltonian \p{AbHam} which can be represented as
\be\label{AbHam1}
H={\overline{\widetilde p}}{}^\alpha\; \left( g^{-1}\right)_\alpha{}^\beta \;{\widetilde p}_\beta+2\; {\cal C}_{U(n)}
\ee
where
\be\label{casU}
{\cal C}_{U(n)} =
 \hJ_\alpha{}^\beta \; \hJ_\beta{}^\alpha -\frac{1}{n+1} \; \hJ_\alpha{}^\alpha \; \hJ_\beta{}^\beta
\ee
is just the $U(n)$ Casimir operator constructed from $U(n)$ currents $\hJ_\beta{}^\alpha$. It is evident that
${\cal C}_{U(n)}$ commutes with all $SU(n+1)$ currents \p{newJsu}, \p{Ab1}. So, from the symmetry point of view one
may consider on the same footing the following Hamiltonian
\be\label{AbHam2}
{\widetilde H}={\overline{\widetilde p}}{}^\alpha\; \left( g^{-1}\right)_\alpha{}^\beta \;{\widetilde p}_\beta+\gamma\; {\cal C}_{U(n)},
\ee
where $\gamma$ is an arbitrary constant. This means that one should use another argument to select the Hamiltonian
with some fixed value of $\gamma$. In principle, one may even consider the reduced system with the fixed value
of the Casimir ${\cal C}_{U(n)}$.

Finally, let us stress that the structure of the currents $\hJ_\beta{}^\alpha$ is completely irrelevant for the
present construction: all that we need is to be sure that these currents span the $U(n)$ algebra \p{inneru}. So, one may
choose these currents to be constructed from some additional isospin degrees of freedom, or one may introduce  new physical bosonic coordinates and their momenta to realize these currents. In this case we will have the system extending standard $\mathbb{CP}^n$ mechanics.

\setcounter{equation}{0}
\section{N=4 Supersymmetry}
In this Section we will extend the consideration from the previous Section to the case of the $N=4$ supersymmetric
extension of $\mathbb{CP}^n$ mechanics.
\subsection{N=4 supersymmetric $\mathbb{CP}^n$ model: free case}
In order to construct the $N=4$ supersymmetric extension of $\mathbb{CP}^n$ mechanics one should introduce $4n$ fermionic variables $\{ \psi_i^\alpha, \bpsi{}^i_\alpha,\; i=1,2\}$ obeying
the following Dirac brackets (together with the previously defined brackets \p{pb1})
\bea\label{PB}
&& \left\{ \psi^\alpha_i, \bpsi_\beta^j\right\} =i \delta_i^j \left( g^{-1}\right)_\beta{}^\alpha, \qquad
\left\{ p_\alpha , \bp^\beta\right\}=
-i \left( g_\alpha{}^\beta g_\mu{}^\nu+g_\alpha{}^\nu g_\mu{}^\beta \right) \bpsi_\nu^i \psi_i^\mu, \nn \\
&& \left\{ p_\alpha, \psi_i^\beta\right\} =-\frac{1}{\left( 1+ z\cdot \bz\right)}
\left[ \bz_\alpha \psi_i^\beta +\delta_\alpha^\beta \,\psi_i^\gamma \bz_\gamma\right], \quad
\left\{ \bp^\alpha, \bpsi^i_\beta\right\} =-\frac{1}{\left( 1+ z\cdot \bz\right)}
\left[ z^\alpha \bpsi^i_\beta +\delta^\alpha_\beta \, z^\gamma \bpsi^i_\gamma \right].
\eea

Now, it is not too hard to check that the supercharges $Q^i, {\overline Q}_i$ have the extremely
simple form \cite{sm1,BN1,BN2}
\be\label{superch}
Q^i = \bp^\alpha \, \bpsi^i_\alpha, \qquad
{\overline Q}_i = \psi^\alpha_i\, p_\alpha.
\ee
They are perfectly anticommute (in virtue of \p{PB}, \p{pb1}) as
\be\label{Q1}
\left\{ Q^i, {\overline Q}_j\right\}= i\delta^i_j H, \qquad
\left\{ Q^i, Q^j\right\} = \left\{{\overline Q}_i, {\overline Q}_j\right\}=0,
\ee
where the Hamiltonian $H$ reads
\footnote{The $su(2)$ indices are raised and
lowered as $A_i=\varepsilon_{ij}A^j, A^i=\varepsilon^{ij} A_j$
with $\varepsilon_{12}=\varepsilon^{21}=1$.}
\be\label{H}
H=\bp^\alpha\, \left( g^{-1}\right)_\alpha{}^\beta \, p_\beta +
\frac{1}{4} \left( g_\mu{}^\alpha g_\rho{}^\sigma+ g_\mu{}^\sigma
g_\rho{}^\alpha \right) \bpsi_{\alpha\, i}\bpsi^i_\sigma\, \psi^{\rho\,j}\psi^\mu_j.
\ee

In the supersymmetric case one may again construct the currents spanning the $su(n+1)$ algebra \p{sunp1}
as\footnote{The explicit expression for generators $J_\alpha{}^\beta$ is not so illuminating and it can be easily
obtained from the definition in \p{sufer}, if needed.}
\bea\label{sufer}
&& R_\alpha= p_\alpha +\bz_\alpha \bz_\beta \bp^\beta-\frac{i}{(1+z\cdot \bz)^2}\left( \bz_\alpha \psi_i^\beta \bpsi{}^i_\beta +\bz_\beta \psi_i^\beta \bpsi{}^i_\alpha -\frac{2}{1+z\cdot \bz} \bz_\alpha \bz_\beta z^\gamma \psi_i^\beta \bpsi{}^i_\gamma \right), \nn\\
&& \bR{}^\alpha=\bp{}^\alpha+z^\alpha z^\beta p_\beta+\frac{i}{(1+z\cdot \bz)^2}\left( z^\alpha \psi_i^\beta \bpsi{}^i_\beta +z^\beta \psi_i^\alpha \bpsi{}^i_\beta -\frac{2}{1+z\cdot \bz} z^\alpha \bz_\beta z^\gamma \psi_i^\beta \bpsi{}^i_\gamma \right),\nn\\
&& J_\alpha{}^\beta = -\frac{i}{2}\left\{ R_\alpha, \bR{}^\beta \right\}.
\eea

One may check that the Hamiltonian \p{H} and the supercharges \p{superch} have vanishing  brackets with the generators of the $su(n+1)$ algebra \p{sufer}:
\be
\left\{ R_\alpha, H\right\}  =  \left\{ \bR{}^\alpha, H\right\}=0, \quad
\left\{ R_\alpha, Q^i\right\}  =  \left\{ \bR{}^\alpha, Q^i\right\}=0,\quad
\left\{ R_\alpha, {\overline Q}_i\right\}  =  \left\{ \bR{}^\alpha, {\overline Q}_i\right\}=0.
\ee
This result is expected. A less expected statement is that the Hamiltonian \p{H} coincides with the Casimir operator of the $su(n+1)$ algebra
\be
H = R_\alpha\; \bR{}^\alpha +2 J_\alpha{}^\beta\; J_\beta{}^\alpha -\frac{2}{n+1} J_\alpha{}^\alpha\; J_\beta{}^\beta ,
\ee
with the currents $\{ R_\alpha, \bR{}^\alpha, J_\alpha{}^\beta\}$ defined in \p{sufer}. Thus, the $N=4$ supersymmetric
$\mathbb{CP}^n$ mechanics has the same symmetry properties as its bosonic core. The only (but crucial) difference is
another realization of the $su(n+1)$ currents \p{sufer} which includes now additional fermionic degrees of freedom.

\subsection{N=4 supersymmetric $\mathbb{CP}^n$ model: interaction}
In this Section, based on the approach presented in the previous Section, we will analyze the symmetry of
the $N=4$ supersymmetric $\mathbb{CP}^n$ model in the background $U(n)$ fields.

In order to introduce the interaction in $N=4$ supersymmetric $\mathbb{CP}^n$ mechanics, following \cite{BKS01}, we
will couple our model with additional  currents $\{ \cR_\alpha, \cbR{}^\alpha, \cJ_\alpha{}^\beta\}$ spanning $SU(1,n)$ groups:
\bea\label{sunm1}
&& \left\{\cR_\alpha, \cbR{}^\beta\right\}  = - 2 i \cJ_\alpha{}^\beta ,\nn \\
&& \left\{\cJ_\alpha{}^\beta, \cR{}_\gamma\right\} = \frac{i}{2} \left( \delta_\gamma^\beta\; \cR_\alpha+\delta_\alpha^\beta\; \cR_\gamma\right) , \quad
\left\{\cJ_\alpha{}^\beta, \cbR{}^\gamma\right\} = -\frac{i}{2} \left( \delta_\alpha^\gamma \; \cbR{}^\beta+\delta_\alpha^\beta\; \cbR{}^\gamma\right), \nn \\
&& \left\{\cJ_\alpha{}^\beta, \cJ_\gamma{}^\delta\right\} = \frac{i}{2} \left( \delta_\gamma^\beta\; \cJ_\alpha{}^\delta-\delta_\alpha^\delta\; \cJ_\gamma{}^\beta\right).
\eea

The model is completely defined by supercharges forming the $N=4$ super Poincare algebra \p{Q1}. Such supercharges have been constructed in \cite{BKS01} as
\be\label{SC1}
Q^i=\bp^\alpha \bpsi^i_\alpha + 2 i\; z^\gamma \cJ_{\gamma}{}^\beta \omega_\beta{}^\alpha \bpsi^i_\alpha +
i\; \psi^{i\,\alpha}e_{\alpha}{}^\beta \cR_\beta, \quad
{\overline Q}_i=\psi^\alpha_i p_\alpha - 2 i\,  \psi^\alpha_i \omega_\alpha{}^\beta \cJ_\beta{}^\gamma \bz_\gamma +
i\; \cbR^\beta e_\beta{}^\alpha \bpsi_{i\,\alpha},
\ee
where $e_\alpha{}^\beta$ and $\omega_\alpha{}^\beta$
are the vielbeins  and $U(n)$-connections  on the $\mathbb{CP}^n\sim SU(n+1)/U(n)$ manifold defined in \p{vb} and \p{con}, correspondingly.

These supercharges are perfectly anticommuting to span the $N=4$ super Poincare algebra \p{Q1}
where the Hamiltonian $H$ now reads
\bea\label{Hfin}
H &=& \left( \bp\, g^{-1}\, p\right) -2 i \left[ \left( \bp\, g^{-1}\, \omega\, \cJ \bz\right) - \left( z\, \cJ\, \omega \, g^{-1}\,p\right)\right]
+\left(\cbR\,\cR\right) +4 \left(z\, \cJ\, \omega\,g^{-1}\,\omega\,\cJ\,\bz\right)\\
&-&  2\, \left (\psi_i\, e\, \cJ \,e \,\bpsi^i \right)
+\frac{1}{4}\, \left(  g_\mu{}^\alpha g_\rho{}^\sigma  + g_\mu{}^\sigma g_\rho{}^\alpha \right)
\bpsi_{\alpha\, i}\bpsi^i_\sigma\, \psi^{\rho\,j}\psi^\mu_j. \nn
\eea

One of the most interesting features of the supercharges \p{SC1}
and Hamiltonian \p{Hfin} is the presence of the full set of
$su(1,n)$ currents \p{sunm1}. If we believe that this Hamiltonian
is $su(n+1)$ invariant, than we have to find the corresponding
currents spanning $su(n+1)$ algebra and having the vanishing
brackets with the Hamiltonian and supercharges. These new extended
$su(n+1)$ currents can contain, besides our bosonic and fermionic
coordinates and momenta, only currents $\cJ_\gamma{}^\delta$
spanning the $u(n)$ algebra. It is clear that the modified
$su(n+1)$ generators cannot contain the currents $\{\cR_\alpha,
\cbR{}^\beta\}$ - it is just impossible to construct new $su(n+1)$
generators from our coordinates, momenta and $su(1,n)$ currents.
This means that the Hamiltonian \p{Hfin} cannot be just a Casimir
operator of some $SU(n+1)$ group. It should have a more
interesting structure. The better understand the structure of the
Hamiltonian \p{Hfin}  one may rewrite the term $\cbR\,\cR$ in
\p{Hfin} as \be\label{gwess1} \cbR{}^\alpha\,\cR_\alpha = {\cal
C}_{su(1,n)} + \left( 2 \cJ_\alpha{}^\beta\; \cJ_\beta{}^\alpha
-\frac{2}{n+1} \cJ_\alpha{}^\alpha\; \cJ_\beta{}^\beta\right) ,
\ee where the $su(1,n)$ Casimir operator ${\cal C}_{su(1,n)}$
reads \be\label{Casm} {\cal C}_{su(1,n)} =
\cbR{}^\alpha\,\cR_\alpha -2 \cJ_\alpha{}^\beta\;
\cJ_\beta{}^\alpha +\frac{2}{n+1} \cJ_\alpha{}^\alpha\;
\cJ_\beta{}^\beta . \ee Thus we see that the unexpected generators
$\{\cR_\alpha, \cbR{}^\beta\}$ belonging to the coset
$SU(1,n)/U(n)$ enter the Hamiltonian only through the $su(1,n)$
Casimir operator ${\cal C}_{su(1,n)}$. The remaining terms depend
only on the $u(n)$ currents $\cJ_\alpha{}^\beta$. Therefore, we
expect that the Hamiltonian \p{Hfin} can be represented as a sum
of two  Casimir operators \be\label{guess2} H= {\widetilde {\cal
C}}_{su(n+1)}+{\cal C}_{su(1,n)}, \ee where the Casimir operator
${\widetilde {\cal C}}_{su(n+1)}$ has to be constructed from our
coordinates, momenta and additional $u(n)$ currents
$\cJ_\gamma{}^\delta$ only. Now we are going to prove this statement.

In order to find these new $su(n+1)$ currents, which commute with the supercharges \p{SC1}, we note that the structure
of new $u(n)$ generators is obvious: there is no other choice besides the direct sum of generators
\be\label{superJ}
{\widetilde J}_\alpha{}^\beta = J_\alpha{}^\beta+\cJ_\alpha{}^\beta.
\ee
Here, the current $J_\alpha{}^\beta$ is the same as in the non-interacting $N=4$ supersymmetric case \p{sufer}. The structure of the coset $SU(n+1)/U(n)$ generators is more involved. In order to find them, one has to consider the most general Ansatz for
these generators compatible with the explicit $U(n)$ symmetry
\bea\label{guess3}
{\widetilde R}_\alpha & =& R_\alpha+i\;f_1\; \cJ_\alpha{}^\beta \bz_\beta +i\; f_2\;\bz_\alpha \;z^\beta \cJ_\beta{}^\gamma \bz_\gamma+i\; f_3\; \bz_\alpha  \cJ_\beta{}^\beta  ,\nn \\
\overline{\widetilde R}{}^\alpha &=& \bR{}^\alpha - i\;f_1\;\; z^\beta \cJ_\beta{}^\alpha{} -i\; f_2\;z^\alpha \;z^\beta \cJ_\beta{}^\gamma \bz_\gamma - i\; f_3\; z^\alpha  \cJ_\beta{}^\beta ,
\eea
where the currents $\{R_\alpha,\bR{}^\alpha\}$ were defined in \p{sufer} and the unknown functions $f_1, f_2, f_3$ depend only on $(z\cdot\bz)$. Then one has to check that: a) these currents span the $su(n+1)$ algebra, b) these currents
commute with the supercharges \p{SC1}.
After a quite lengthy calculation, one may find these function to be
\be\label{f}
f_1 = \frac{2}{\left(1+\sqrt{1+ z \cdot \bz}\right)}, \quad f_2 = \frac{1}{\left(1+\sqrt{1+ z \cdot \bz}\right)^2},
\quad f_3=0,
\ee
and therefore
\bea\label{sufin}
&& {\widetilde R}_\alpha = R_\alpha+\frac{2 i}{\left(1+\sqrt{1+ z \cdot \bz}\right)}\; \cJ_\alpha{}^\beta \bz_\beta +\frac{ i}{\left(1+\sqrt{1+ z \cdot \bz}\right)^2}\;\bz_\alpha \;z^\beta \cJ_\beta{}^\gamma \bz_\gamma ,\nn \\
&& \overline{\widetilde R}{}^\alpha = \bR{}^\alpha - \frac{2 i}{\left(1+\sqrt{1+ z \cdot \bz}\right)}\; z^\beta \cJ_\beta{}^\alpha{} -\frac{ i}{\left(1+\sqrt{1+ z \cdot \bz}\right)^2}\;z^\alpha \;z^\beta \cJ_\beta{}^\gamma \bz_\gamma ,\nn \\
&& {\widetilde J}_\alpha{}^\beta = -\frac{i}{2}\left\{ {\widetilde R}_\alpha, \overline{\widetilde R}{}^\beta \right\}=J_\alpha{}^\beta+\cJ_\alpha{}^\beta.
\eea

All these generators commute with the supercharges $\{Q^i, {\overline Q}_i\}$ and, therefore, with the Hamiltonian
\p{Hfin}, while the Hamiltonian itself \p{Hfin} has the suggested structure \p{guess2}. Thus, we conclude that $N=4$ supersymmetric $\mathbb{CP}^n$ mechanics in the background $U(n)$ fields possesses $su(n+1)$ symmetry generated by the
currents \p{sufin}. Similarly to the bosonic cases, the structure of the $su(1,n)$ currents $\{ \cR_\alpha, \cbR{}^\alpha, \cJ_\alpha{}^\beta\}$ \p{sunm1} is irrelevant for our construction. One may also consider a reduced version of the system by fixing the value of the Casimir operator ${\cal C}_{su(1,n)}$, which clearly commutes with everything.

\setcounter{equation}{0}
\section{Conclusion}
In the present paper we have proved the $SU(n+1)$ invariance of a $N=4$ supersymmetric extension of mechanics describing
the motion of a particle over the $\mathbb{CP}^n$ manifold in the presence of background $U(n)$ gauge fields. We have explicitly constructed the corresponding $su(n+1)$ currents which commute with the supercharges and showed that the Hamiltonian of
the system can be represented as a direct sum of two Casimir operators. One Casimir operator, on the $SU(n+1)$ group, contains our bosonic and fermionic coordinates and momenta together with additional $U(n)$ currents constructed from isospin degrees of freedom, while the second one, on the $SU(1,n)$ group, contains isospin degrees of freedom only.

Proving the $SU(n+1)$ invariance of a $N=4$ supersymmetric mechanics describing
the motion of a charged particle over $\mathbb{CP}^n$ manifold in the presence of background $U(n)$ gauge fields \cite{BKS01} is
crucial for the possible application of this model to the analysis of the role that the additional fermionic variables play in the quantum Hall effect on $\mathbb{CP}^n$. Surely, this is one of the most interesting immediate applications of our results.

The approach we used in this paper, in order to visualize the symmetry of the supersymmetric $\mathbb{CP}^n$ model, is based mainly on its interpretation as a sigma model on the coset $SU(n+1)/U(n)$. In this respect, viewing the
$\mathbb{CP}^n$ manifold as the coset $Sp(k+1)/U(1)\times Sp(k)$ \cite{add1} opens a way not only to turning on the background fields, living on $U(1)\times Sp(k)$, but also for the construction of the corresponding supersymmetric extensions. Using
$U(1)\times Sp(k)$ background fields instead of those living on the $U(n)$ group could bring some new features into
QHE on $\mathbb{CP}^n$ manifold. This will be described in more detail elsewhere.

\section*{Acknowledgements}
We are indebted to Armen Nersessian  for valuable discussions.

S.K. and A.S. thank the INFN-Laboratori Nazionali di Frascati, where this work was completed, for
warm hospitality.

This work was partly supported by Volkswagen Foundation grant I/84 496, by RFBR grants  11-02-01335-a,
 11-02-90445-Ukr, 12-02-00517-a, as well as by the ERC Advanced Grant no. 226455, \textit{``Supersymmetry, Quantum Gravity and Gauge Fields''} (\textit{SUPERFIELDS}).

\end{document}